\documentclass{JHEP3}

\usepackage{amssymb}
\usepackage{amsmath}
\usepackage{graphicx}

\newcommand{\mpl}{m_{\mathrm{pl}}}
\newcommand{\ud}{\mathrm{d}}
\newcommand{\pd}{\partial}
\def\f {{\phi}}
\def\ff{{\phi\phi}}
\def\fff{{\phi\phi\phi}}
\def\Pr{{P_{\cal R}}}
\def\calO{{\cal O}}
\def\pmpc{{\rm Mpc^{-1}}}

\title{A Brachistochrone Approach to Reconstruct the Inflaton Potential}

\author{ Daniel Wohns$^1$\footnote{dfw9@cornell.edu}~,
Jiajun Xu$^2$\footnote{jxu47@wisc.edu} and S.-H. Henry Tye$^1$\footnote{sht5@cornell.edu} \\
{\em $^1$Laboratory of Elementary Particle Physics, Cornell University, \\
 \hspace{0.5ex} Ithaca, NY 14850, USA \\
$^2$Department of Physics, University of Wisconsin-Madison, \\
 \hspace{0.5ex} Madison, WI 53705, USA}}

\abstract{
We propose a new way to implement an inflationary prior to a cosmological dataset that incorporates the inflationary observables at arbitrary order. This approach employs an exponential form for the Hubble parameter $H(\phi)$ without taking the slow-roll approximation. At lowest non-trivial order, this $H(\phi)$ has the unique property that it is the solution to the brachistochrone problem for inflation.
}

\begin{document}

\maketitle

\section{Introduction}

In inflationary universe, we can use the evolution of the inflaton as a clock.
For a single scalar field with canonical kinetic term, the inflationary Hubble parameter
can be expressed as a function of $\phi$ only:
\begin{equation}
H(\phi, \dot\phi) \rightarrow H(\phi)
\end{equation}
Since we need an almost flat inflaton potential for enough e-folds of inflation, we may choose to start with a truncated expression \cite{inflation_flow}, for some $M$,
\begin{equation}
\label{Hflat}
H(\phi) = H_0 (1 + b_1 \f + b_2 \f^2/2 + \dots + b_M \f^M/M!)
\end{equation}
where $\phi$ is in Planck units and $b_i$'s are dimensionless. The coefficients $b_i$ are expected to be small, so the truncation to $M+1$ terms, with a relatively small $M$,
should yield a very good approximation, especially for the region where CMB and other cosmological data are available to constrain them. Most efforts to implement an inflationary
prior to a cosmological data set along this line use $H(\phi)$ of this form \cite{Hansen:2001eu, Kinney:2003uw, Peiris:2006sj}.  An alternative approach uses the inflationary flow formalism to reconstruct models without assuming slow roll \cite{Powell:2007gu}.

One might worry that different parameterizations would lead to different constraints on the inflationary parameters.  It was shown in \cite{Hamann:2008pb} that third-order Taylor expansions of $H(\phi)$ and $H^2(\phi)$ with a prior of sufficient e-folds yield the same constraints on the observable window of inflation.

The conditions for a reasonable trial $H(\phi)$ are: first, it is close to what data indicate, so the Taylor expansion needs only a few terms; second, the expression is easy to manipulate when calculating observables.
Third, it will be nice if the leading term has a simple physical interpretation. In the above case, the leading term is simply a flat potential yielding an unlimited number of e-folds. $H(\phi)$ of Eq.(\ref{Hflat}) is just a perturbed version of a flat inflaton potential.

We like to ask if there is another expression for $H(\phi)$ which has these three properties. Furthermore, it will be really useful if one does not have to take the slow-roll approximation. Here we claim that there is such an expression, which in addition has a conceptual underpinning behind it. Consider
\begin{equation}
\label{form1}
H^{(M)}(\phi) = H_0 \exp(a_1\f + a_2\f^2/2 + a_3 \f^3/3! + \dots + a_M \f^M/M!) ~,
\end{equation}
here $H^{(0)}(\phi)=H_0$ corresponds to a flat potential. What is particularly interesting about this exponential form is that, when it is truncated to the linear term in the exponent, i.e.,
\begin{equation}
\label{brach}
H^{(1)}(\phi) = H_0\exp(a_1 \phi) = H_0 \exp(\sqrt{\epsilon /2}\, \phi)
\end{equation}
it is simply the solution to the brachistochrone problem for inflation without taking the slow-roll approximation. That is, away from the slow-roll approximation, $H^{(1)}(\phi)$ yields the minimum number of e-folds (fastest path) for a fixed drop in $H$ over a fixed field range. So, any deviation from this path will yield more e-folds. In this sense, it is the opposite of the flat potential.

The exponential form of $H(\phi)$ was also mentioned in an early paper \cite{Liddle:1993ch} by Andrew Liddle, as a way to construct arbitrary inflaton potential. Here,  we motivate the form Eq.(\ref{form1}) from entirely new perspective, i.e. the brachistochrone problem for inflation and how to deviate away from the brachistochrone solution. In this sense, we have given the exponential of $H(\phi)$ an interesting physical interpretation.

Furthermore, it is quite amazing that, for small $\epsilon$, $H^{(1)}(\phi)$ alone (i.e., without higher terms in Eq.(\ref{form1})) can yield an inflationary scenario within
experimental bounds, implying a low (small $M$, say $M \gtrsim 2$) $H^{(M)}(\phi)$ will do very well already.  Models that are close to the brachistochrone solution typically require a spectral index that is within a few percent of unity.  Lastly, the relations between the slow-roll parameters and the coefficients $a_i$ in $H^{(M)}(\phi)$ of Eq.(\ref{form1}) are very simple.



\section{The Minimal E-folds in Canonical Single Field Inflation}

The dynamics of a single canonical scalar field during inflation can be described by the Hamilton-Jacobi formalism \cite{Salopek:1990jq}, where we write every quantity as a function of the scalar field $\phi$, i.e., we choose $\phi$ as the ``clock''. We will use the notation $H_\f \equiv \ud{H}/\ud{\f}$, $H_\ff \equiv \ud^2 H/ \ud\f^2$, etc.
\begin{eqnarray}
3 H^2 &=& \frac{1}{2} \dot{\phi}^2 + V(\phi) ~, \label{HJ1} \\
-\dot\phi &=& 2 H_\f(\phi) \label{HJ2} ~.
\end{eqnarray}
We set $\mpl^{-2} \equiv 8\pi G = 1$ throughout the paper. Here we do not consider violation of the Null Energy Condition, so $\dot{H} < 0$.  We choose the convention $\dot{\phi} < 0$ so that $\ud{H}/\ud{\phi} > 0$.

The $\epsilon$ and $\eta$ parameters in this formalism are given by
\begin{eqnarray}
\epsilon &\equiv& -\frac{\dot H}{H^2} \;=\; 2 \left( \frac{H_\f}{H} \right)^2  ~, \label{eq_eps} \\
\eta &\equiv& \frac{\dot\epsilon}{H\epsilon} \;=\; 4 \left[ \left( \frac{H_\f}{H} \right)^2 - \frac{H_\ff}{H} \right] ~. \label{eq_eta}
\end{eqnarray}
The number of e-folds is given by
\begin{equation}
\label{Nef}
N_e = \int \frac{H}{\dot\phi} \ud\phi = -\frac{1}{2} \int \frac{H}{H_\f} \ud\phi ~.
\end{equation}

Consider the case with fixed initial Hubble scale $H_0 \equiv H(\phi=0)$ and final Hubble scale $H_f \equiv H(-\Delta\phi)$ over a fixed finite range $\Delta\phi$ of $\phi$. We can ask what kind of function $H(\phi)$ gives the minimum number of e-folds between these two fixed boundary points. Starting with Eq.(\ref{Nef}), one can write down the Euler-Lagrangian equation,
\begin{equation}
\frac{\pd}{\pd H} \left( \frac{H}{H_\f} \right) = \frac{\pd}{\pd \phi} \frac{\pd}{\pd H_\f} \left( \frac{H}{H_\f} \right)
\end{equation}
which gives
\begin{equation}
\frac{H_\ff}{H} - \left(\frac{H_\f}{H}\right)^2 = 0 \label{euler_eq}
\end{equation}
Comparing with the definition of $\eta$ in Eq.(\ref{eq_eta}), the Euler-Lagrangian equation is equivalent to saying that $\eta = 0$ for the trajectory with minimum e-folds.

The solution to Eq.(\ref{euler_eq}) is given by
\begin{equation}
H(\phi) = H_0 \exp \left(a_1 \phi \right)
\end{equation}
where the integration constant $a_1$ is related to the $\epsilon$ parameter as $$\epsilon = 2a_1^2$$ For inflation, we require
\begin{eqnarray}
0 < \epsilon < 1  \quad \Rightarrow \quad 0 < a_1 < \frac{\sqrt{2}}{2} ~.
\end{eqnarray}
The lower bound $a_1 > 0$ arises because $\ud{H}/\ud{\phi} > 0$.



From $H(\phi)$ above, we can reconstruct the potential through Eq.(\ref{HJ1}) and Eq.(\ref{HJ2}),
\begin{eqnarray}
V(\phi) = \left( 3 - \epsilon \right)
H_0^2 \, \exp \left( \sqrt{2\epsilon}\, \f \right) ~.
\end{eqnarray}

In the limit $\epsilon \ll 1$, the potential becomes linear
\begin{equation}
V_S(\phi) \approx 3 H_0^2 \left( 1 + \sqrt{2\epsilon } \, \f \right) ~,
\end{equation}
which agrees with previous analysis in the slow-roll scenario with large damping \cite{Tye:2009ff}
\footnote{
Note that, to get the correct power spectrum index,  we have to keep the $\phi^2$ term in $V_S(\phi)$.
Using Eq.(\ref{dlnk}) one obtains, $n_s -1 =\ud\ln\Pr/\ud\ln k = -(2 \epsilon +\eta) / (1 - \epsilon) \simeq  -2 \epsilon -\eta \simeq -6 \epsilon +2 \eta_{SR}$ where the last equality is taken in the slow-roll approximation.}.

One can now calculate the value of the minimum number of e-folds
\begin{equation}
N_e^{\mathrm{min}} = \frac{-1}{2} \int \ud\phi \frac{H(\phi)}{H_\f(\phi)} = \frac{(\Delta\f)^2}{2\ln (H_0/H_f)} ~.
\end{equation}

Since the minimal e-fold path corresponds to $\eta = 0$, we can turn on the $\eta$ parameter to characterize the deviation away from the minimal e-fold path. With non-zero $\eta$, we need to solve
\begin{equation}
\left(\frac{H_\f}{H}\right)^2  - \frac{H_\ff}{H} = \frac{\eta(\phi)}{4} \label{eq_Heta} ~.
\end{equation}
The solution can be formally written as
\begin{equation}
H(\phi) = H_0 \exp \left[ a_1' \f - \frac{1}{4} \int_0^\phi \ud{\phi_1} \int_0^{\phi_1} \ud{\phi_2}
\,\eta(\phi_2) \right] ~.
\label{sol_Heta}
\end{equation}
We can fix the integration constant $a_1'$ by matching $H(-\Delta\phi) = H_f$. Therefore
\begin{equation}
a_1'= -\frac{\ln (H_f/H_0)}{\Delta\f} - \frac{1}{4 \Delta\f} \int_0^{-\Delta\phi} \ud{\phi_1} \int_0^{\phi_1} \ud{\phi_2} \,\eta(\phi_2) ~.
\end{equation}
Now the number of e-folds is
\begin{equation}
N_e = \frac{1}{2} \int \frac{H}{H_\f} \ud{\phi}
= \frac{1}{2} \int^0_{-\Delta\f}  \left(a_1' -\frac{1}{4} \int_0^\f \eta(\f_1) \,\ud{\f_1} \right)^{-1} \ud{\phi}
\end{equation}

In principle, one can consider an arbitrary function of $\eta(\phi)$, as long as $|\eta(\phi)| < 1$. Here we will illustrate using a constant $\eta$, so that the solution Eq.(\ref{sol_Heta}) can be simplified,
\begin{eqnarray}
H(\phi) &=& H_0 \exp \left(a_1' \phi - \frac{\eta}{8} \phi^2 \right) \\
a_1' &=&  -\frac{\ln (H_f/H_0)}{\Delta \phi} - \frac{\eta}{8}\Delta\phi \\
N_e &=& \frac{2}{\eta} \ln \left[ \frac{\ln(H_f/H_0) - \frac{1}{8}\eta(\Delta\phi)^2}{\ln(H_f/H_0) + \frac{1}{8}\eta(\Delta\phi)^2} \right]
\end{eqnarray}
In the small $\eta$ expansion, we can write
\begin{equation}
N_e = \frac{(\Delta\phi)^2}{2\ln(H_0/H_f)} + \frac{(\Delta\f)^6 \eta^2}{384 [\ln (H_0/H_f)]^3} + {\cal O}(\eta^4) + \dots ~,
\end{equation}
which further verifies that $\eta \to 0$, $N_e$ approaches the minimum.


Instead of the ansatz (\ref{form1}), we can also use the following ansatz
\begin{equation}
\label{form2}
H(\phi) = H_0 \, e^{a_1\f}\left(1 + b_1\f + b_2\f^2/2 +. . .\right)
\end{equation}
where
\begin{equation}
\epsilon=2 (a_1+b_1)^2
\end{equation}
This interpolates between the two limiting starting functional forms for the inflaton potential.
If we want to increase by one parameter only, we can simply set $b_1=0$,
\begin{equation}
H(\phi) = H_0 \, e^{a_1 \f}\left(1 +  b_2 \f^2/2  \right)
\end{equation}
Now
\begin{equation}
\epsilon=2 (a_1+b_1)^2=2a_1^2 ~,  \quad \quad  \eta= 4 (b_1^2 - b_2)=-4b_2 ~.
\end{equation}
Starting from $H^{(1)}(\phi)$, it is unclear which is the best ansatz to generalize it, Eq.(\ref{form1}) or Eq.(\ref{form2}).  It will depend on future data.

\section{The Power Spectrum}\label{pr}

We have seen that the form of $H(\phi)$ we proposed has some theoretical underpinning: $H^{(1)}(\phi)$ is the form that gives the minimum number of e-folds, and $H^{(2)}(\phi)$ can be interpreted as deforming away from the minimal e-fold trajectory using a constant $\eta$ parameter. Going beyond $H^{(1)}$, it is not clear what is the best way to generalize the form of $H(\phi)$. At least we have seen two possibilities, Eq.(\ref{form1}) and Eq.(\ref{form2}). In this section, we will try to argue that the form Eq.(\ref{form1}) is more favorable in the sense that inflationary observables can be more simply expressed in terms of its parameters. We will see that the tensor and scalar power spectrum, their spectral index, running of spectral index, etc., can all be written entirely in terms of $\ln H$ and its $\phi$ derivatives, which makes the exponential form of Eq.(\ref{form1}) extremely convenient for computing observables.

Let us first look at the scalar power spectrum
\begin{equation}
\Pr(k) = \frac{H^2}{8 \pi^2 \epsilon} = \frac{H^2}{16\pi^2 (\ln H)_\f^2} \Big|_{\f = 0} ~,
\end{equation}
where, without loss of generality, we have chosen $\f = 0$ to correspond to the moment when the wave vector $k$ crosses the horizon.

Furthermore, we notice that the $\epsilon$ and $\eta$ parameters can be written in terms of $\ln H$ and its derivatives
\begin{eqnarray}
\epsilon = 2\, (\ln H)_\f^2 ~, \quad \eta = -4\, (\ln H)_\ff ~.
\end{eqnarray}
We also have the derivative
\begin{equation}\label{dlnk}
\frac{\ud}{\ud \ln k} = \frac{\ud \phi}{\ud \ln k}\frac{\ud}{\ud \phi} = \frac{-2 (\ln H)_\f}{1-\epsilon} \frac{\ud}{\ud \phi}
\end{equation}

We see that $\ln{H}$ appears everywhere, which suggests that if we Taylor expand $\ln{H}$ instead of $H$, i.e., using the exponential form in Eq.(\ref{form1}), the resulting $\ud
\ln k$ derivatives will be very easy to compute. We now show explicit computations of $n_s$, $\ud n_s/\ud \ln k$, $r$ and $n_t$ starting with $H(\phi)$ of Eq.(\ref{form1}).
\begin{eqnarray}
\frac{\ud \ln \Pr}{\ud \ln k}\Big|_{\f = 0} &=& \frac{-2 (\ln H)_\f}{1-\epsilon} \frac{\ud}{\ud \f}
\left[ 2\ln H - 2 \ln \left( -(\ln H)_\f\right) \right] \Big|_{\f = 0} \nonumber \\
&=& \frac{4\, (\ln H)_\ff - 4 (\ln H)_\f^2}{ 1 - 2 (\ln H)_\f^2} \Big|_{\f = 0} \nonumber \\
&=& \frac{4a_2 -4a_1^2}{1 - 2a_1^2}
\end{eqnarray}
so the scalar spectral index is
\begin{eqnarray}
\label{ns}
n_s = 1 + \frac{4a_2 - 4a_1^2}{1 - 2a_1^2} \approx 1 + 4a_2 - 4a_1^2
\end{eqnarray}
To compare with the $\epsilon$ and $\eta$ expansion, simply replace $\epsilon = 2a_1^2$, $\eta = -4a_2$, and we get the usual result $n_s = 1 - 2\epsilon - \eta$. The advantage of writing $H(\phi)$ in exponential form is that by derivatives of $\ln H$, one can extract the coefficients $a_i$'s  more transparently.

Comparing to the usual $\epsilon$ and $\eta$ expansion, here the order of expansion can be tracked by counting the total number of $\phi$ derivatives. So $a_1^2$ and $a_2$ are of the same order, since they measure the two $\phi$ derivative terms. The parameter $a_3$ is of higher order, since it measures the term with one more derivative of $\phi$.

To further appreciate the advantage of the exponential form, let us compute the running of the spectral index. Up to four derivatives, we get
\begin{eqnarray}\label{exp_running}
\frac{\ud n_s}{\ud \ln k} \approx \frac{-2 (\ln H)_\f}{1-\epsilon} \frac{\ud}{\ud \f}
\left[ 4\, (\ln H)_\ff - 4 (\ln H)_\f^2 \right] \Big|_{\f = 0} = \frac{-8a_1a_3 + 16a_1^2a_2}{1-2a_1^2}
\end{eqnarray}

In a more careful analysis including $\calO(\epsilon)$ corrections in the power spectrum,
\[
\Pr = \frac{H^2}{16\pi^2 (\ln H)_\f^2} \left[ 1 - (C+1)\epsilon - \frac{1}{2}C \eta \right]^2 ~, \quad C=-2+\ln2+\gamma
\]
where $\gamma$ is the Euler-Mascheroni constant, we get up to four derivatives,
\begin{eqnarray}
n_s &\approx& 1 + \frac{4a_2 - 4a_1^2}{1 - 2a_1^2} + \frac{-4 (\ln H)_\f}{1-\epsilon} \Big[ - 4(C+1)(\ln H)_\f (\ln H)_\ff
+ 2C (\ln H)_\fff  \Big] \nonumber \\
&=& 1 + \frac{4a_2 - 4a_1^2}{1 - 2a_1^2} + \frac{16(C+1)a_1^2 a_2 - 8C a_1 a_3}{1 - 2a_1^2} \label{ns_2}
\end{eqnarray}
and up to six derivatives
\begin{eqnarray} \label{exp_running_2}
\frac{\ud n_s}{\ud \ln k} & \approx &
\frac{-2(\ln H)_\f}{1 - 2(\ln H)_\f^2} \frac{\ud }{\ud \phi}
\Big[
4 (\ln H)_\ff - 4(\ln H)_\f^2 + 8(\ln H)_\ff(\ln H)_\f^2 - 8(\ln H)_\f^4  \nonumber \\
&& + 16(C+1) (\ln H)_\f^2 (\ln H)_\ff - 8C (\ln H)_\f (\ln H)_\fff
\Big] \nonumber \\
&=& \frac{16 a_1^2 a_2 - 8a_1a_3}{1-2a_1^2}
+ \frac{1}{1-2a_1^2}\Big[64a_1^4a_2 - (32C+48) a_1^3 a_3 \nonumber \\
&& - (64C+96) a_1^2a_2^2 + 16C a_1^2 a_4 + 16C a_1 a_2 a_3 \Big]
\end{eqnarray}

Similarly, for the tensor power spectrum
\[
P_T = \frac{2H^2}{\pi^2} \left[1-(C+1)\epsilon\right]^2
\]
we get the tensor to scalar ratio
\begin{eqnarray}
\label{r}
r &\approx& 32 (\ln H)_\f^2 \left( 1 + C \eta \right) \nonumber \\
&=& 32 a_1^2 (1 - 4 C a_2) ~,
\end{eqnarray}
and the tensor spectral index
\begin{eqnarray}
\label{n_t}
n_t  
&\approx& \frac{-2 (\ln H)_\f}{1-2 (\ln H)_\f^2} \left[ 2 (\ln H)_\f - 8 (C+1) (\ln H)_\f (\ln H)_\ff \right] \nonumber \\
&=& \frac{-4a_1^2 + 16 (C+1)a_1^2 a_2 }{1-2a_1^2}
\end{eqnarray}

Our approach can compute higher derivatives $\ud^n n_s/(\ud \ln k)^n$ ($n > 3$) very efficiently, as all the derivatives of $\ln H$ are tied to its Taylor expansion coefficients
directly. The computational effort does not increase much with taking more derivatives, and new parameters $a_4, a_5, \dots, a_M$ enter in a systematic way. The total number of
$\phi$ derivatives in each term can be used as order counting in the expansion. Whereas in the usual approach, the higher derivatives of the expansion parameters $\epsilon$, $\eta$, $\xi$ are tedious to work out. Clearly, the exponential form of $H(\phi)$ has its advantage in computing power spectrum observables.


\section{Data Constraints}

We use Markov Chain Monte Carlo methods to fit WMAP7 data, using modified versions of CAMB \cite{Lewis:1999bs} and COSMOMC \cite{Lewis:2002ah}.  The independent parameters we
choose are $\{a_1, a_2, a_3, a_4 \}$. All other cosmological parameters are fixed at the central values in the $\Lambda$CDM fit of the WMAP7 data in Ref. \cite{LAMBDA}. We parametrize the scalar and tensor primordial power spectrum by
\begin{eqnarray}
\Pr &=& A_s \left( \frac{k}{k_0} \right)^{n_s-1 + \frac{1}{2} (\ud n_s/ \ud \ln k) \ln (k/k_0) } \\
P_T &=& r A_s \left( \frac{k}{k_0} \right)^{n_T}
\end{eqnarray}
The relation between $\{ n_s,\, \ud n_s /\ud \ln k,\, r \}$ and $\{ a_1, a_2, a_3, a_4 \}$ are worked out in Sec.\ref{pr}. The scalar amplitude is fixed at $A_s = 2.23 \times 10^{-9}$ by adjusting the parameter $H_0$, i.e. given $a_1$ and $A_s$, we set $H_0/\mpl$ through the relation
\[
A_s  = \frac{H_0^2}{16\pi^2 a_1^2 \mpl^2} ~.
\]
Since we fix $A_s$, $a_1$ is correlated with $H/\mpl$. For inflation, we require $0 < a_1 < \sqrt{2}/2$. In our data analysis, we put a lower bound on $a_1$ by hand, i.e. we searched the parameter space with $10^{-3} \lesssim a_1 < \sqrt{2}/2$.

\begin{figure}
\begin{center}
\includegraphics[width=15cm]{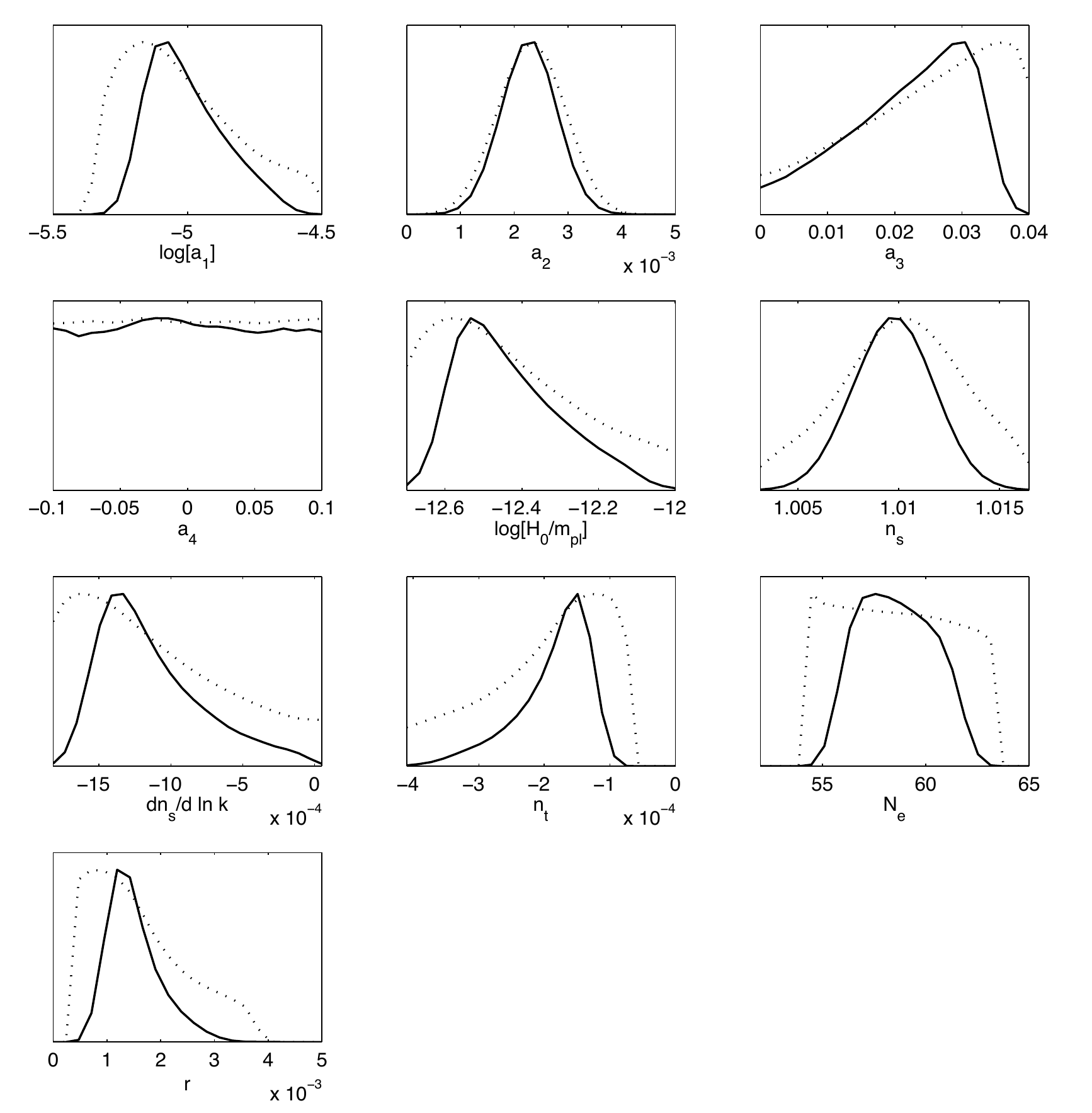}
\end{center}
\caption{One-dimensional marginalized constraints of parameters from WMAP7. $\{a_1, a_2, a_3, a_4\}$ are free parameters and the rest are derived using equations in Sec. \protect\ref{pr}. Solid lines show the full marginalized probability density on a linear scale. Dashed lines show the mean likelihood of samples on a linear scale.} \label{1dfig}
\end{figure}

\begin{figure}
\begin{center}
\includegraphics[width=15cm]{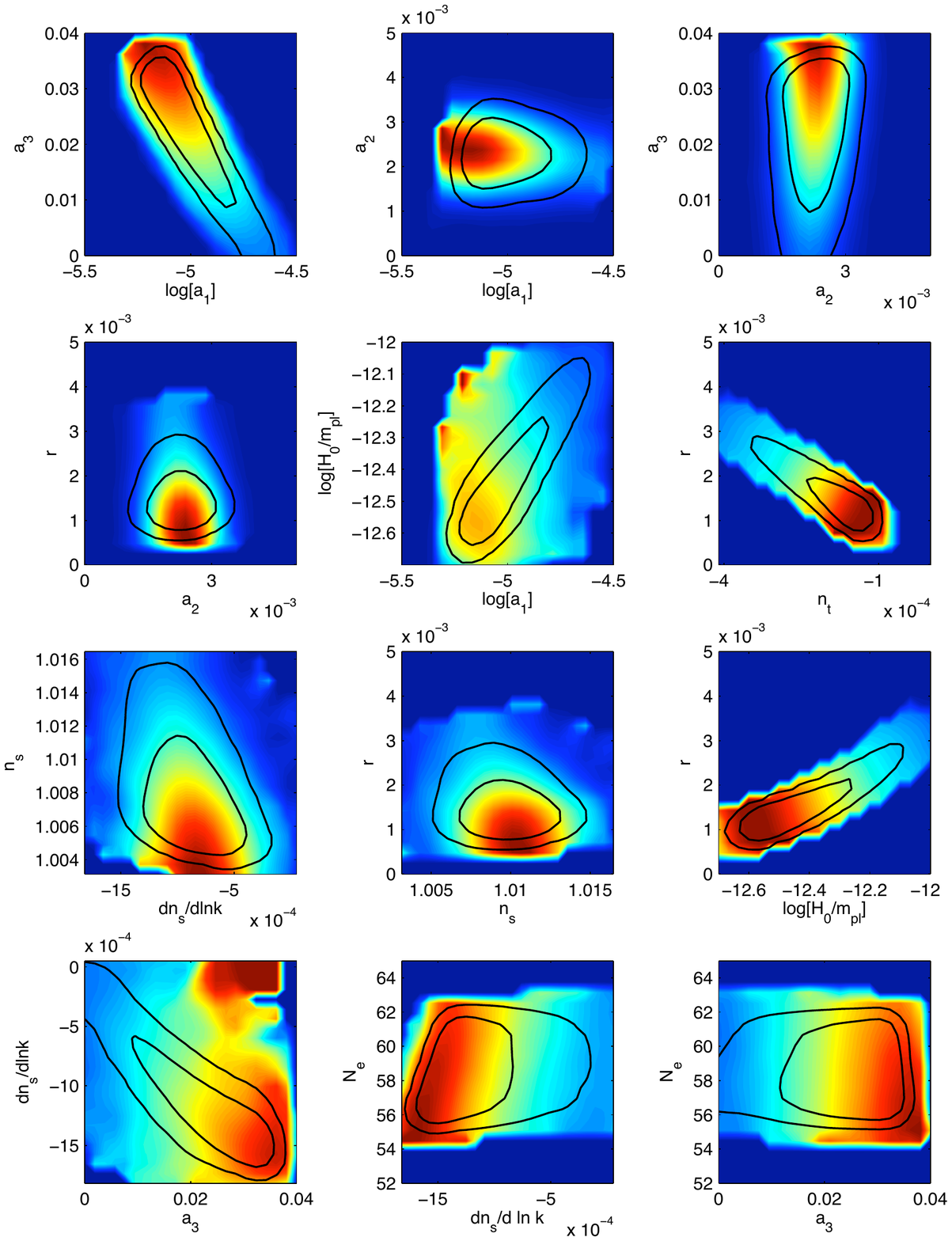}
\end{center}
\caption{Two-dimensional marginalized distributions from WMAP7 data. The contours represent 68\% and 95\% confidence levels. The color represents mean likelihood of the samples with red being high likelihood, followed by yellow.} \label{2dfig}
\end{figure}

The theoretical priors we impose are
\begin{itemize}

\item Perturbative expansion of $\ln (H)$: We require that higher-order terms in the Taylor expansion of $\ln (H)$ are smaller than leading-order terms, that is
\begin{eqnarray}
\label{pertexp}
a_1 \phi > a_2 \phi^2 ~, \quad a_1 \phi > a_3 \phi^3 ~, \quad \dots
\end{eqnarray}

\item Monotonic function for $H(\phi)$: During inflation, $\dot\phi$ cannot change sign, which implies that $H(\phi)$ is a monotonic function. In our convention, we require $H_\f
> 0$. For the cubic truncation of $\ln (H)$, $a_4 = 0$, $H_\f >0$ is equivalent to the following constraints on $a_1$, $a_2$ and $a_3$
\begin{eqnarray} \label{constr_Hp}
a_3 > 0 ~, \quad a_2^2 - 2a_1 a_3 < 0 ~.
\end{eqnarray}
If $a_4 \ne 0$ but small, the above constraints will be relaxed, but we do not expect qualitative change of the parameter space. As we will see, data currently are not sensitive to $a_4$. In our analysis, we impose the same constraint (\ref{constr_Hp}) even if $a_4 \ne 0$.


\item The end of inflation: We assume there are two ways to end inflation in our approach: (1) Inflation ends when $\epsilon$ grows and becomes larger than unity. (2) Although
$\epsilon$ does not grow larger than one during inflation, when the inflaton travels for $\Delta\phi \sim 1$, higher-order terms in $H(\phi)$ becomes important; they may change the form of the potential to shut off inflation.

One can estimate how much the inflaton field travels before $\epsilon$ grows to one. This gives
\begin{eqnarray}
\Delta \phi = \frac{- a_2 - \sqrt{a_2^2 + \sqrt{2}a_3 - 2a_1 a_3} }{a_3} ~.
\end{eqnarray}
Note that with $a_3 > 0$ and $a_1 < \sqrt{2}/2$, we have $a_2^2 + \sqrt{2}a_3 - 2a_1 a_3 > 0$, so $\Delta \phi$ is always well defined.

Therefore, inflation ends with $\epsilon = 1$ if
\begin{eqnarray}
|\Delta \phi| = \frac{a_2 + \sqrt{a_2^2 + \sqrt{2}a_3 - 2a_1 a_3} }{a_3} < 1 ~.
\end{eqnarray}
Otherwise, inflation ends when $\Delta \phi = -1$.

\item Number of e-folds: We require that after the fiducial scale $k_0 = 0.002\, \pmpc$ has left the horizon, inflation will last for $N_e$ e-folds. Since we are only interested in large scale observations, we constrain $N_e$ in a small window
\begin{eqnarray}
N_{\rm pivot} - 3 < N_e < N_{\rm pivot} + 3 ~,
\end{eqnarray}
while $N_{\rm pivot}$ is correlated with $H/\mpl$ through
\begin{eqnarray}
N_{\rm pivot} \approx  65 - \ln\left(\frac{k_{\rm pivot}}{0.002}\right) + 0.5 \ln \left(\frac{H_0}{\mpl}\right)
\end{eqnarray}
By imposing the lower bound on $N_e$, we discard models which do not yield enough e-folds to solve the horizon problem. The upper bound on $N_e$ makes sure that our parameters $\{a_1, a_2, a_3, a_4\}$, which are defined at $\phi=0$, correspond to scales within the observable window of WMAP data.
\end{itemize}

Fig.\ref{1dfig} and Fig.\ref{2dfig} show constraint on our parameters $\{a_1, a_2, a_3, a_4\}$ from WMAP7 data. We have run eight MCMC chains and they converge with the Gelman and Rubin statistics $R-1 \lesssim 0.01$. Since we only use the data analysis for illustration, we have not run the MCMC for longer time to achieve better convergence.

We see that data constrain the parameter $a_1$ to be around $a_1 \approx 0.006$, corresponding to $\epsilon \sim 10^{-4}$. This is consistent, since we are essentially dealing with small field inflation with $\Delta \phi \lesssim 1$. The total number of e-folds are $N_e \sim 1/\sqrt{\epsilon} \sim 10^2$.

The lower bound on $a_1$ may come as a surprise at first sight. One may think that given a reasonable value of $a_1$, we can always change the value of $H/\mpl$ to fit the scalar
amplitude, so there should not be a lower bound on $a_1$ as shown in Fig.\ref{1dfig}. In fact, the lower bound comes from our theoretical prior $H_\f > 0$, i.e.
Eq.(\ref{constr_Hp}), which leads to $a_1 > a_2^2/(2a_3)$ for given $a_2$ and $a_3$. One may increase $a_3$ to allow smaller value of $a_1$, but on the other hand, our
perturbative truncation requires that $a_3 \phi^3 < a_1 \phi$, so $a_3$ cannot be increased without limit. In summary, the constraint on $a_1$ mainly comes from the set of theoretical priors we applied in the data analysis, in particular our choice $\Delta \phi = -1$.

As $\epsilon \sim 10^{-4}$, its contribution to $n_s$ is negligible. So $n_s$ is determined mostly by $a_2$. That is why $a_2$ becomes the best constrained parameter among the four. Currently, WMAP7 data do not constrain $a_3$ and $a_4$ very well, which suggests that the quadratic truncation $H(\phi)$ in Eq.(\ref{form1}) already provides a very good fit of data.


Our analysis gives a slightly blue spectral index with a small negative running parameter. This is consistent with the WMAP7 analysis \cite{Komatsu:2010fb}. When we allow the spectral index to run, the central value of $n_s$ shifts towards the blue-tilted side. However, one should not take seriously our result of a nonzero $\ud n_s / \ud \ln k$. This is most likely an artifact due to fixing other cosmological parameters. For example, Ref.\cite{Peiris:2006sj} has shown that the value of total matter density $\Omega_M$ is strongly correlated with the running of spectral index. We expect that when marginalizing over all $\Lambda$CDM parameters, the running of spectral index will likely become insignificant.

At fixed order, our theoretical priors impose an upper bound on $|n_s - 1|$.  At second order this bound is, using Eqs. (\ref{ns}) and (\ref{pertexp}):
\begin{equation}
|n_s - 1| < \frac{4 a_1 (\Delta \phi)^{-1} + 4 a_1^2}{1-2a_1^2} = \frac{2\sqrt{2 \epsilon}(\Delta \phi)^{-1} + 2 \epsilon}{1-\epsilon}
\end{equation}
At third order this bound is:
\begin{equation}
|n_s - 1| < \frac{2\sqrt{2\epsilon} (\Delta \phi)^{-1} + (2-4C (\Delta \phi)^{-2}) \epsilon +
4\sqrt{2}(C+1)\epsilon^{3/2} (\Delta \phi)^{-1}}{1-\epsilon}
\end{equation}
where $C \approx -0.7$.  This bound gets progressively weaker at higher order.  For $\Delta \phi = 1$ and $\epsilon \ll 1$ at low order,
\begin{equation}
\label{nsbound}
|n_s - 1| \lesssim 2\sqrt{2 \epsilon} .
\end{equation}
For $\epsilon \approx 10^{-4}$, Eq. (\ref{nsbound}) implies $|n_s - 1| \lesssim 0.03$.
A small perturbation of the brachistochrone solution (\ref{brach}) with $\Delta \phi = 1$ is only a good fit to data when Eq. (\ref{nsbound}) is satisfied.  Note that in deriving this bound, we assumed that the leading order term in (\ref{form1})is dominant and that the form (\ref{form1}) is valid throughout the final sixty e-folds of inflation.  Relaxing one or both of these assumptions would eliminate our theoretical prior and allow the spectral index to deviate further from unity, but render the brachistochrone interpretation that originally motivated our approach inapplicable.

The constraints on $r$ are driven primarily by the constraints on $a_1$. That means the reported bounds on tensor to scalar ratio is primarily from model, i.e. the inflationary prior, not from data. The bound on $r$ from data is much weaker according to the WMAP analysis \cite{Komatsu:2010fb}. In our analysis, we focus on the set of models with $\Delta \phi < 1$, which will limit the level of tensor mode according to the Lyth Bound $\Delta\phi / \Delta N_e \sim \sqrt{r/8}$. Our method can also be applied to cases with $\Delta \phi > 1$, but then one have to worry about the truncation of higher order terms in the expansion of $\ln H$. We will leave such an analysis for future work.

It will be very interesting to include more data sets in our analysis, such as ACBAR, Large Scale Structure and PLANCK data. With a wider span of scales, we may be able to achieve
better constraints on $a_3$ and $a_4$. In addition, PLANCK will provide us with a tighter bound on tensor mode, which will help to further constrain $a_1$. Furthermore, given our
exponential form of $H(\phi)$ and the polynomial form in Ref.\cite{inflation_flow, Peiris:2006sj}, it will be interesting to see which functional form is favored by data, and to see if single-field models can be distinguished experimentally from more general models \cite{Easson:2010zy, Easson:2010uw}. Since PLANCK and other data sets will become available soon, we leave the full data analysis and Bayesian model comparison for future work.

\section{Conclusions and Remarks}

In this paper, we have proposed a new functional form of $H(\phi)$ as an effective way to implement inflationary priors in cosmological data analysis. This new form of $H(\phi)$
entails writing $H$ in the exponential form. At the lowest order, with the exponent linear in $\phi$, it is the solution to the brachistochrone problem for inflation, which
corresponds to the minimal number of e-folds for a fixed drop in $H$ over a fixed field range. Higher-order terms can be included to deviate the inflaton path away from the brachistochrone solution.

In addition to the theoretical underpinning, the exponential form of $H(\phi)$ also provides an efficient way to compute power spectrum observables. If one Taylor expands the function $\ln (H(\phi))$, the expansion coefficients are natural parameters to express the observables, such as $n_s$, $\ud n_s /\ud \ln k$, $r$ and $n_t$. Higher-order derivatives of the power spectrum entails higher-order expansion coefficients of $\ln (H(\phi))$, and the computation is more straightforward than the usual $\epsilon$ and $\eta$ parametrization.

We have also performed MCMC analysis to illustrate how observation data can constrain the expansion coefficients of $\ln (H(\phi))$. The WMAP7 data constrains the two leading coefficients very well, but is not quite sensitive to the coefficients of the cubic and quartic terms. We expect including more data on different scales will improve the constraints. With the upcoming precision measurements from Cosmic Microwave Background and Large Scale Structure, we hope our proposal will offer an efficient way to reconstruct the inflaton potential from data.

In the actual implementation of an inflationary prior to a cosmological dataset, one does not need to use the same $H(\phi)$ for the whole range of sixty e-folds of inflation. The actual inflaton potential can be more complicated; in fact, it can be multi-field. In these more general situations, one can use a set of piecewise exponential segments of the form (\ref{form1}) instead.
In this case, the constraint of sixty or more e-folds imposed in the above analysis may be relaxed.

\vspace{0.6cm}

\noindent {\bf Acknowledgments}

\vspace{0.3cm}

DW thanks N. Agarwal for technical assistance.  DW and HT are supported by the National Science Foundation under grant
PHY-0355005.  JX was supported in part by a DOE grant DE-FG-02-95ER40896, a Cottrell Scholar Award from Research Corporation, and a Vilas Associate Award.

\end{document}